\documentclass[prl,twocolumn,showpacs,preprintnumbers,superscriptaddress]{revtex4}
\usepackage{graphicx}
\usepackage{color}
\usepackage{dcolumn}
\usepackage{bm}
\usepackage{graphics}
\usepackage{revsymb}
\usepackage[caption=false]{subfig}
\usepackage{amsbsy}
\usepackage{amsmath}
\usepackage{amsfonts}
\usepackage{amsthm}
\usepackage{float}
\usepackage{natbib}
\usepackage{multirow}
\usepackage{hyperref}

\theoremstyle{plain}

\theoremstyle{definition}
\begin{document}

\title{Sequential Quantum Secret Sharing in a Noisy Environment aided with Weak Measurements}
\author{M. Ray}
\author{S. Chatterjee}
\affiliation{Centre for Computational Natural Sciences and Bioinformatics,}
\author{I. Chakrabarty}
\affiliation{Center for Security Theory and Algorithmic Research \\
International Institute of Information Technology, Gachibowli, Hyderabad-500032, Andhra Pradesh, India.}

\begin{abstract}

In this work we give a $(n,n)$-threshold protocol for sequential secret sharing of quantum information for the first time. 
By sequential secret sharing we refer to a situation where the dealer is not having all the secrets at the same time, 
at the beginning of the protocol; however if the dealer wishes to share secrets at subsequent phases she/he can realize it 
with the help of our protocol. First of all we present our protocol for three parties and later we generalize it for the situation where we have more $(n>3)$ parties.  
Interestingly, we show that our protocol of sequential secret sharing requires less amount of quantum as well as classical resource as compared to the situation wherein existing protocols 
are repeatedly used. Further in a much more realistic situation, we consider the sharing of qubits through two kinds of noisy channels, 
namely the phase damping channel (PDC) and the amplitude damping channel (ADC). 
When we carry out the sequential secret sharing in the presence of noise we observe that the fidelity of secret sharing 
at the $k^{th}$ iteration is independent of the effect of noise at the $(k-1)^{th}$ iteration. In case of ADC we have seen that the average fidelity of secret sharing drops
down to $\frac{1}{2}$ which is equivalent to a random guess of the quantum secret. Interestingly, we find that  by applying weak measurements one can enhance the average 
fidelity. This increase of the average fidelity can be achieved with certain trade off  with the success probability of the weak measurements. 

\end{abstract}

\maketitle

\section{Introduction}

For a long time quantum entanglement ("spooky action at a distance") \cite{einstein} was only of philosophical interest 
until people had found out various ways of utilizing it as a resource for information processing protocols like 
quantum teleportation \cite{bennett2}, quantum super-dense coding, entanglement broadcasting and quantum cryptography \cite{gisin,hillery}.

In quantum cryptography when we work with quantum systems, entanglement probes to be an useful resource in carrying
out various protocols. Quantum secret sharing is no exception to it. In a nut shell, secret sharing refers to a situation 
where the sender shares a secret message between other parties in such a way that none of them can reveal the secret 
without the collaboration of others.

In quantum secret sharing (QSS) \cite{hillery, cleve99} generally we deal with the problem of sharing of both classical as 
well as quantum secrets. This is done by using quantum resources like entangled states; mostly pure entangled state. 
These protocols include various attempts on tightening the security in presence of eaves droppers. It is not necessary that 
we always require three qubit entangled state as a resource for the most simplest secret sharing protocols. Karlsson 
et.al.\cite{karlsson} showed that similar quantum secret sharing protocols using bipartite pure entangled states also exist. 
However in general research was carried out mainly investigating the concept of quantum secret sharing using tripartite pure 
entangled states and  multipartite states like graph states \cite{bandyopadhyay,bagherinezhad,lance,gordon,zheng, markham,markham08} 
as resources. More precisely when we talk about multiqubit secret sharing, we generally talk about a situation where the dealer wants to send multiple secrets; is a QSS scheme to various reconstructors. Apart from classical and quantum secret sharing, protocols have been given to share semi-quantum secrets using entangled states as the resource \cite{li}.

Recently in the reference \cite{satya}, it was shown that quantum secret sharing is possible even with bipartite 
two qubit mixed states formed due to the transmission of qubits through noisy environment. In a realistic situation, 
the secret sharing of classical or quantum information will involve the transmission of qubits through noisy environment. 
As a result of which the resource state will become a mixed-state and the secret sharing will no longer be a deterministic one.
In reference \cite{indranil}, authors proposed a protocol for secret sharing of classical information in the presence of such noise.
In different works like \cite{tittel, schmid, schmid1, bogdanski}, it has been shown that quantum secret sharing is not only a mere 
theoretical concept, but also an experimental possibility.

In another piece of work authors have investigated the revocation of quantum secret back to the dealer and routing
quantum information to different receivers in a network \cite{sazim-2013}. In the long run quantum secret sharing is also an important area 
to study in various quantum networks (QNet) as in recent times researchers have seen that both classical as well as quantum information 
can be transferred elegantly through quantum networks \cite{sazim-2012}.

It is quite well-known that environmental interactions incurring loss of fidelity of the shared secret 
is an ubiquitous process and unless controlled using well-formulated schemes can reduce a shared secret 
in a pure state to one in maximally mixed state during the reconstruction phase. Hence it becomes imperative 
to devise schemes of improving the fidelity of shared secret. In references \cite{Koashi-1999, Kim-2009, Lee-2011, Kim-2012}, 
authors have suggested that one can apply weak measurements \cite{Aharonov-1988} to protect the fidelity of 
quantum states subjected to decoherence through an amplitude damping channel (ADC) and this technique is even 
shown to be practically implementable. Recently in \cite{Pramanik-2013}, authors have shown how efficiently 
one can employ the technique of weak measurement with post selection and its reversal to improve the fidelity 
of teleportation through an ADC. They also show how to exploit the power of post selection \cite{Aharonov-1988} 
in weak measurements to work with sub-ensemble of the initial states and hence reduce the suppression of decoherence 
of the transmitted qubits.

In this work we have given a protocol for sequential multi qubit secret sharing. 
This sequential secret sharing
is useful when the dealer is not having all the secrets at the beginning of the protocol and wishes to send the secrets at subsequent stages. 
By our protocol one can achieve sequential secret sharing using lesser number of qubits and classical bits of communication than any other existing quantum secret sharing protocols when performed repeatedly.
We have also considered a realistic situation when the qubits are transferred from the dealer to other parties through noisy 
quantum channel. In that case we have seen that the fidelity of secret sharing at the $k^{th}$ iteration is independent of the 
effect of noise at the $(k-1)^{th}$ iteration. Finally we employ the technique of weak measurement with post selection 
and its reversal to improve the fidelity of the shared secret under the effect of ADC. For any given input state parameters 
and strength of decoherence channel, we find out for what values of strength parameters of weak measurement and reverse 
weak measurement will help one to gain maximum fidelity of reconstructing the shared secret. We also show how the success 
probability of such an improvement technique decreases with increase in strength of weak measurement and that of the decoherence channel.

\section{Quantum Sequential Secret Sharing in the absence of noise}

In this section, we present a protocol by which Alice (the dealer) will securely share the secrets 
$\psi_{1},\psi_{2},....\psi_{m}$ among $n$ parties at various stages of the protocol. 
By this we refer to a situation where the dealer is not having all the secrets at the beginning of the protocol. 
So he/she shares them at subsequent stages depending on the availability of the secrets. 

\subsection{Sequential Secret Sharing with three Parties}
In this subsection we begin with three parties, where Alice is the dealer and Bob and Charlie are the receivers.
Alice wants to share the secrets $|\psi_{i}\rangle=\alpha_{i}|0\left\rangle _{1}\right.+\beta_{i}|1\left\rangle _{1}\right.$, 
(where $|\alpha_i^2|+|\beta_i^2|=1$ and $i$ is from 1 to $m$) at the $i^{th}$ iteration of the protocol, to both Bob and Charlie. 
However to do so Alice needs to share a quantum resource with the other two parties.
In order to prepare this resource state Alice starts with a Bell state, 
say $|\phi^{+}\rangle=\frac{1}{\sqrt{2}}\left\{|00\left\rangle _{23}+|11\left\rangle _{23}\right.\right.\right\}$. 
The combined system of the Bell state and the secret 
$|\psi_{1}\rangle=\alpha_{1}|0\left\rangle +\beta_{1}|1\left\rangle \right.\right.$, available to her at the first iteration 
is given by, 
\begin{eqnarray} 
|\psi_{1}\rangle|\phi^{+}\rangle=\frac{1}{\sqrt{2}}\left(\alpha_{1}|000\rangle_{123}+\alpha_{1}|011\rangle_{123}\right) \nonumber\\
+\frac{1}{\sqrt{2}}\left(\beta_{1}|100\rangle_{123}+\beta_{1}|111\rangle_{123}\right). \label{eq:combsys}
\end{eqnarray}
Now Alice carries out a XOR operation between qubits 1 and 2 to obtain the three qubit entangled state 
\begin{eqnarray}
|\Psi_{in}\rangle=
\frac{1}{\sqrt{2}}(\alpha_{1}|000\rangle _{123}+\alpha_{1}|011\rangle _{123}\nonumber\\
+\beta_{1}|110\rangle _{123}+\beta_{1}|101\rangle _{123}). \label{eq:inpstate}
\end{eqnarray}
Then Alice keeps the qubit 2 with herself and sends the remaining qubits 1 and 3 to Charlie and Bob respectively. Here we consider 
the transfer of qubits to take place in an idealistic situation where the medium of transfer is free of noise.  
The three qubit entangled state shared between them can be re-written as
\begin{eqnarray}
|\Psi_{in}\rangle=\frac{1}{\sqrt{2}}[|0\rangle_{2}
\{|+\rangle_{1}(\alpha_{1}|0\rangle _{3}+\beta_{1}|1\rangle _{3})+
\nonumber\\
|-\rangle_{1}(\alpha_{1}|0\rangle _{3}-\beta_{1}|1\rangle _{3})\}+
|1\rangle{}_{2}\{|+\rangle_{1}(\alpha_{1}|1\rangle _{3}+\nonumber\\
\beta_{1}|0\rangle _{3})+|-\rangle_{1}(\alpha_{1}|1\rangle _{3}-
\beta_{1}|0\rangle _{3})\}].  \label{eq:final3state}
\end{eqnarray}
After that Alice measures her qubit in the computational basis $\{|0\rangle,|1\rangle\}$ and Charlie
measures his qubit in $\{|+\rangle,|-\rangle\}$ basis. At this point, Alice does not tell either Bob or Charlie
about her measurement outcomes.  This implies that the single-qubit density matrices of both Bob's and Charlie's 
qubits are $(1/2)I$, where $I$ is the $2\times 2$ identity matrix. Thus at this stage of the protocol neither Bob nor Charlie
has any information about Alice's qubit. Unlike other protocols, it is interesting to note that in our protocol we are 
carrying out single qubit measurements at two different locations instead of two qubit Bell measurement. 
Once the measurement is over from Alice's side the secret is shared between other two parties, Bob and Charlie.
However neither of them can reveal the secret without the collaboration of the other party. This completes the sharing of Alice's secret.
Now if Bob wants to reveal the secret, then both Alice and Charlie have to send their measurement outcomes to Bob in form of one 
classical bit. Depending on the outcomes of Charlie and Alice, Bob can reveal the secret by applying appropriate set of unitary 
transformations. This set of unitary transformations is shown in the following (TABLE \ref{tbl:operators}).
\begin{table}[h]
\caption{\bf Sharing of quantum secrets \label{tbl:operators}} 
\begin{center}
\begin{tabular}{|c|c|}
\hline Outcomes of Alice and Charlie   & Unitary Transformations \\
\hline $|0\rangle,|+\rangle$  & $I$  \\
\hline $|0\rangle,|-\rangle$  & $\sigma_{z}$  \\
\hline $|1\rangle,|+\rangle$  & $\sigma_{x}$  \\
\hline $|1\rangle,|-\rangle$  & $-i\:\sigma_{y}$  \\
\hline
\end{tabular}
\end{center}
\end{table} 

However if Alice wishes to carry out further secret sharing, Charlie needs 
to send his qubit to Alice. On receiving the qubit from Charlie, Alice applies the unitary transformation to convert 
it into $|0\rangle$ state. In addition to that at each iteration Alice needs the state 
$\frac{1}{\sqrt{2}}\left(\left|0\right\rangle +\left|1\right\rangle \right)$. 
With this single qubit state and the state obtained from Charlie she applies XOR operation to re-create a Bell
pair $\left|\phi^{+}\right\rangle =\frac{1}{\sqrt{2}}\left\{ \left|00\right\rangle _{23}+\left|11\right\rangle _{23}\right\}$. 
Then she follows the same steps to carry out further secret sharing. The key point of the protocol is that 
in order to carry out sequential secret sharing, at the end of each iteration Alice needs the supply  
of the qubit $\frac{1}{\sqrt{2}}\left(\left|0\right\rangle +\left|1\right\rangle \right)$ along with the qubit transferred by Charlie. 

\subsection{Sequential Secret Sharing with Multi-parties}

In this subsection, we extend our protocol for a situation where we have more than three parties and consider a much more generalized
condition having $n$ parties other than the dealer herself. 
The protocol for multi party secret sharing is mainly a natural extension of our protocol for three parties. However the resource state 
required for secret sharing in this case is an $n+1$ qubit pure entangled state given by, 
\begin{eqnarray}
|\Psi_{in}^{n+1}\rangle=\frac{1}{\sqrt{2}}(\alpha_{i}|000..0\rangle _{123..(n+1)}+\alpha_{i}|011..1\rangle _{123..(n+1)}\nonumber\\
+\beta_{i}|110..0\rangle _{123..(n+1)}+\beta_{i}|101..1\rangle _{123..(n+1)}),  \label{eq:multi}
\end{eqnarray}
where $\alpha_{i}$ and $\beta_{i}$ are the input state parameters. 

This state is obtained by Alice after performing a XOR operation on the secret available to her at the first iteration with any one of the 
qubits of the n qubit GHZ state ($|GHZ\left\rangle \right._n=\frac{1}{\sqrt{2}}[{|000...0\rangle+|111...1\rangle}]$).  
Alice measures the second qubit in  $\{|0\rangle,|1\rangle\}$ basis 
and distributes the qubit such that the party who is supposed to reconstruct the secret (say Bob) is given the 1st qubit. 
In the reconstruction phase all the other parties have to measure their qubits in  $\{|+\rangle,|-\rangle\}$ basis and 
convey their results to Bob. 
If Alice wishes to go for further secret sharing the qubits are to be returned back to her by others except Bob. 
She applies the appropriate unitary transformation to convert them 
to $\left|0\right\rangle$ state and uses the qubit $\frac{1}{\sqrt{2}}\left(\left|0\right\rangle +\left|1\right\rangle \right)$  
along with these $n-1$ qubits to perform a chained XOR. This results in a resource state given by,  

\begin{eqnarray}
\frac{1}{\sqrt{2}}\left\{ \underset{n}{\left|\underbrace{000...0}\right\rangle}+\underset{n}{\left|\underbrace{100...0}\right
\rangle}\right\} \nonumber\\
\overset{XOR}{\longrightarrow}\frac{1}{\sqrt{2}}\left\{ \underset{n}{\left|\underbrace{000...0}\right\rangle}+
\underset{n}{\left|\underbrace{111...1}\right\rangle}\right\}. \label{eq:multiXOR}
\end{eqnarray}

\subsection*{Resource Requirements} 

\begin{figure}[!ht]
\begin{flushleft}
\[
\begin{array}{cc} 
\includegraphics[height=5cm,width=9.5cm]{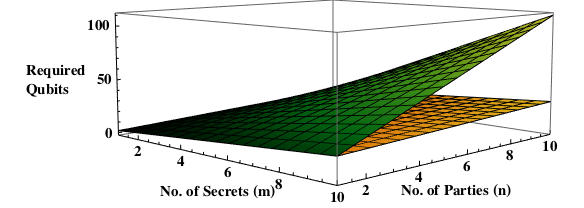}
\end{array}
\]
\caption{\noindent \scriptsize Comparison of qubit requirements of our protocol (orange plane) with the existing ones (greenish-yellow plane). \label{fig:comparison}} 
\end{flushleft}
\end{figure}

We have $n$ parties and $m$ secrets to be shared. If we use the existing protocols repeatedly then we would need $m(n + 1)$ number (which grows quadratically) of qubits \cite{hillery}. However, with the above protocol we need $n+1$ qubits for the first round and only two more in each of the subsequent rounds. Therefore we only need  $n+1+2(m-1)=2m+n-1$ (which grows linearly) qubits. The difference between the resource requirements of the existing protocols and our protocol is $m(n+1)-(2m+n-1)=(m-1)(n-1) \geq 0,\:\: \forall\: m, n  \geq 1$. This shows that the resource requirement for our protocol is strictly less than that of the existing protocols when used repeatedly for sequential sharing of quantum secrets. In the Fig.~\ref{fig:comparison}, we compare the requirement of resource qubits in our protocol with the existing ones. The sharp rise of the greenish-yellow plane describing the qubit requirement with existing protocols demonstrates the rapid growth in the amount of required qubits with increase in values of $m$ as well as $n$, while the low rise of the orange plane describing the required number of qubits in our case shows that the requirement of qubits here is strictly lesser. Moreover, the rapid increase in separation between the two planes with higher values of $m$ and $n$ explicitly expresses the utility of our protocol over the existing ones for performing sequential secret sharing in terms of resource requirements. Apart from this our protocol also requires lesser number of classical bits to be communicated to the secret re-constructor. It takes $nm$ bits of classical communication in the earlier case, whereas it takes $(n-1)m$ bits in our protocol.
\section{Quantum Sequential Secret Sharing in the presence of noise}
In this section, we consider a more realistic situation where three parties share a three qubit mixed entangled state instead 
of a three qubit pure entangled state as a resource. Similarly, here also, we start with a pure entangled state $|\Psi_{in}\rangle$, 
initially with all the qubits with Alice. Now  Alice keeps the second qubit with herself and sends the remaining qubits 
(1 \& 3) to Charlie and Bob. She sends them through a noisy quantum channel. Since practically no channel is completely noise-free,
such an analysis in the absence of noise is imperative. We study our protocol under the effects of both phase damping channel(PDC) and amplitude damping 
channel(ADC). 

\subsection{Transmission of qubits through a PDC}
In the first subsection, we study the effects of PDC on the secret sharing fidelity when the qubits are sent through this channel. 
The action of a PDC are given by the set of three Kraus operators, namely,  
$K_{0}=\sqrt{1-q}I$, $K_{1}=\sqrt{q}\left|0\right\rangle \left\langle 0\right|$, and 
$K_{2}=\sqrt{q}\left|1\right\rangle \left\langle 1\right|$; where $q(0\leq q \leq1)$ is the channel strength \cite{Nielsen-2002}.
The action of the Kraus operators, describing the PDC, on the two qubits results in a three qubit mixed entangled state, 
$\rho_{out}^{}$ are as follows,



\begin{equation}
\rho_{out}=\underset{}{\sum_i}\underset{}{\sum_j}[(K_{i}\otimes I\otimes K_{j})\rho^{in}(K_{i}^{\text{\dag}}\otimes I^{\text{\dag}}\otimes K_{j}^{\text{\dag}})] \label{eq:noiserow}
\end{equation}


where $i,j\,\epsilon\,\{0,1,2\}$. 
\textcolor{black}{Here the Kraus operators acts on the two qubits sent to Bob and Charlie by Alice. These qubits are the 
ones which are subjected to environmental interaction and thus undergo phase damping.} 
In the reconstruction phase, Alice measures her qubit in $\{|0\rangle,|1\rangle\}$ basis and Charlie measures in $\{|+\rangle,|-\rangle\}$ basis. 
After these measurements Bob's state collapses to one of the four states, $\rho_{out}^{a,b}$ (with $a\in \{0,1\}$ representing
Alice's measurement outcome and $b\in \{+,-\}$, representing Charlie's measurement outcome). These four states are given by,

\begin{eqnarray}
&&\rho_{out}^{0,+}=\alpha^{2}\left|0\right\rangle \left\langle 0\right|+\beta^{2}\left|1\right\rangle \left\langle 1\right|+{}\nonumber\\&&(1-q)^{2}\alpha\beta\left|1\right\rangle \left\langle 0\right|+(1-q)^{2}\alpha\beta\left|0\right\rangle \left\langle 1\right|{}\nonumber\\&&
\rho_{out}^{0,-}=\alpha^{2}\left|0\right\rangle \left\langle 0\right|+\beta^{2}\left|1\right\rangle \left\langle 1\right|-{}\nonumber\\&&(1-q)^{2}\alpha\beta\left|1\right\rangle \left\langle 0\right|-(1-q)^{2}\alpha\beta\left|0\right\rangle \left\langle 1\right|{}\nonumber\\&&
\rho_{out}^{1,+}=\alpha^{2}\left|1\right\rangle \left\langle 1\right|+\beta^{2}\left|0\right\rangle \left\langle 0\right|+{}\nonumber\\&&(1-q)^{2}\alpha\beta\left|1\right\rangle \left\langle 0\right|+(1-q)^{2}\alpha\beta\left|0\right\rangle \left\langle 1\right|{}\nonumber\\&&
\rho_{out}^{1,-}=\alpha^{2}\left|1\right\rangle \left\langle 1\right|+\beta^{2}\left|0\right\rangle \left\langle 0\right|-{}\nonumber\\&&(1-q)^{2}\alpha\beta\left|1\right\rangle \left\langle 0\right|-(1-q)^{2}\alpha\beta\left|0\right\rangle \left\langle 1\right|. \label{eq:fourstatesPDC}
\end{eqnarray}

Since the channel through which the qubits are sent is a noisy channel, the qubit obtained by Bob is no longer the desired qubit but a single qubit mixed state $\rho_{out}^{a,b}$. At this point we define the fidelity of quantum secret sharing as the overlap between the original secret and the nearest possible state obtained during the reconstruction phase after optimizing over all possible complex unitaries. The expression for the secret sharing fidelity given by, 
\begin{eqnarray}
\underset{U}{Maximize}\,\left\langle \psi\left|U\rho_{out}^{a,b}U^{\dagger}\right|\psi\right\rangle\
\end{eqnarray}
where $U$ represents complex unitaries. 

In our case, after doing optimization over all possible complex unitary matrices, we find that the unitaries that will take the state obtained at Bob's side are the same set of unitaries that are given in TABLE \ref{tbl:operators} and the fidelity of secret sharing is given by,
\begin{eqnarray}
F_{PD}=\alpha^{4}+2(1-q)^{2}\alpha^{2}\beta^{2}+\beta^{4}. \label{eq:FPDC}
\end{eqnarray} 

\begin{figure}[!ht]
\begin{center} 
\[
\begin{array}{cc} 
\includegraphics[height=5cm,width=8cm]{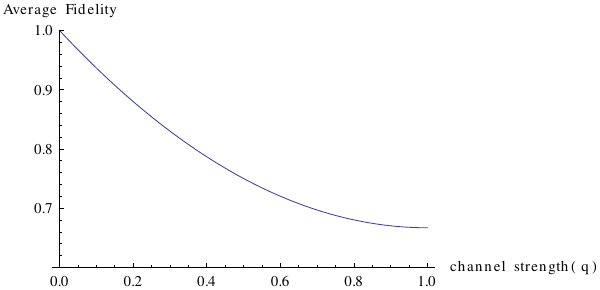}
\end{array}
\]
\caption{\noindent \scriptsize Variation of Average Fidelity of sequential secret sharing with the PDC parameter($q$). \label{fig:fid_pdc}} 
\end{center}
\end{figure}

The average fidelity obtained after averaging over all input parameter $\alpha^2$ is given by, 
\begin{equation}
\overline{F}_{PD}=1-\frac{2q}{3}+\frac{q^2}{3}. \label{eq:avgFPDC}
\end{equation}
In FIG. ~\ref{fig:fid_pdc} we plot the average fidelity ($\overline{F}_{PD}$) against the channel strength ($q$). It is evident from the figure itself 
that when we have no noise in the channel, that is $q=0$, we have $\overline{F}_{PD}=1$. This ensures the fact that in the absence of noise 
we can always do perfect secret sharing. However, for maximum noise in the phase damping channel ($q=1$) we have the value of the average fidelity 
to be $\frac{2}{3}$. This result is analogous to teleportation with classical channel and implies that for full phase damping case these channels 
become classical: no phase information can be transmitted, but a classical bit can be reliably communicated \cite{massar-1995}.  

\subsection{Transmission of qubits through an ADC}
In this subsection, we investigate the process of secret sharing when the qubits are sent through amplitude damping channel (ADC). 

The action of an ADC are given by the set of two Kraus operators, namely,  
$K_{0}=\left|0\right\rangle \left\langle 0\right|+\sqrt{1-p}\left|1\right\rangle \left\langle 1\right|$, and
$K_{1}=\sqrt{p}\left|0\right\rangle \left\langle 1\right|$; where $p(0 \leq p \leq 1)$ is the channel strength \cite{Nielsen-2002}.
The action of the Kraus operators, describing the ADC, on the two qubits (sent to Bob and Charlie) 
results in a three qubit mixed entangled state, $\rho_{out}^{}$. 

The four states corresponding to the measurement outcomes are given by,

\begin{eqnarray}
&&\rho_{out}^{0,+}=(\alpha^{2}+p\beta^2)\left|0\right\rangle \left\langle 0\right|+(1-p)\beta^{2}\left|1\right\rangle \left\langle 1\right|+{}\nonumber\\&&(1-p)\alpha\beta\left|1\right\rangle \left\langle 0\right|+(1-p)\alpha\beta\left|0\right\rangle \left\langle 1\right|{}\nonumber\\&&
\rho_{out}^{0,-}=(\alpha^{2}+p\beta^2)\left|0\right\rangle \left\langle 0\right|+(1-p)\beta^{2}\left|1\right\rangle \left\langle 1\right|-{}\nonumber\\&&(1-p)\alpha\beta\left|1\right\rangle \left\langle 0\right|-(1-p)\alpha\beta\left|0\right\rangle \left\langle 1\right|{}\nonumber\\&&
\rho_{out}^{1,+}=(p\alpha^{2}+\beta^2)\left|0\right\rangle \left\langle 0\right|+(1-p)\alpha^{2}\left|1\right\rangle \left\langle 1\right|+{}\nonumber\\&&(1-p)\alpha\beta\left|1\right\rangle \left\langle 0\right|+(1-p)\alpha\beta\left|0\right\rangle \left\langle 1\right|{}\nonumber\\&&
\rho_{out}^{1,-}=(p\alpha^{2}+\beta^2)\left|0\right\rangle \left\langle 0\right|+(1-p)\alpha^{2}\left|1\right\rangle \left\langle 1\right|-{}\nonumber\\&&(1-p)\alpha\beta\left|1\right\rangle \left\langle 0\right|-(1-p)\alpha\beta\left|0\right\rangle \left\langle 1\right| \label{eq:fourstatesADC}
\end{eqnarray}

Depending on measurement outcome of Alice and Charlie, Bob applies the
corresponding unitary transformations according to the ones given by TABLE \ref{tbl:operators}. 

Proceeding similar to the above case of PDC, the fidelity of quantum secret sharing is obtained by optimizing 
over all possible complex unitaries and is given by 
\begin{equation}
\overline{F}_{AD}=\alpha^{2}+(1-p)\beta^{2}. \label{eq:FADC}
\end{equation}

Unlike the case of PDC, here we cannot convert all the four $\rho_{out}^{a,b}$ exactly to $\rho_{out}^{0,+}$ using unitary operations. 
However we can convert them pairwise. 
By applying the unitary operations in TABLE \ref{tbl:operators}, we end up with two different fidelity expressions. 
Both these expressions, when averaged over all possible input states, give rise to the same average fidelity for the protocol. 
The average fidelity obtained after averaging over all input parameter $\alpha^2$ is given by 
\begin{equation}
\overline{F}_{AD}=1-\frac{p}{2} \label{eq:avgFADC}
\end{equation}

It is evident from the Eq.(~\ref{eq:FADC}) that when $p=0$ we have $\overline{F}_{AD}=1$, ensuring the fact that in the 
absence of noise we can always do perfect secret sharing. For $p=1$, in presence of full amplitude 
damping strength no information can be transmitted over the channel and the fidelity is $\frac{1}{2}$ corresponding 
to random guessing. The output of the channel is always $|0\rangle$ regardless of the input. 
  
For both above channels, we observe that in our protocol the fidelity of secret sharing in subsequent iterations remains the same and is 
independent of the number of iterations. This is because at each and every iteration there is reusability of Charlie's qubit. 
Though Charlie's qubit is affected by noise and will also get further affected when
he sends back his qubit to Alice, this noise is not going to play a part in the subsequent iterations. This is because Alice on
getting back the qubit from Charlie measures it in the computational basis $\{|0\rangle,|1\rangle\}$ and transforms it accordingly
to the state $|0\rangle$ by applying the appropriate unitary transformation.
\section{Improving the fidelity of the shared secret using weak measurements}

In the previous section we have noticed that in a realistic situation the shared qubits are subjected to 
decoherence due to the presence of noise. This environmental interaction results in inevitable loss of fidelity 
of obtaining the shared secret. The reconstructed secret under noise becomes a mixed state and under certain conditions, for 
full phase damping channel, the average fidelity drops to $\frac{2}{3}$ where the phase information is lost and it becomes 
equivalent to transmission over a classical channel. The case is even worse for full amplitude damping channel, 
the average fidelity drops to $\frac{1}{2}$ where it conveys of no more information than what can be obtained 
by employing random guess. Thus, there arises an exigency to improve the fidelity of reconstructing the shared secret.

The authors in the references \cite{Koashi-1999, Kim-2009, Lee-2011, Kim-2012}, have shown that one can reduce the effect of amplitude 
damping decoherence with the help of weak measurement and reverse quantum measurement (WMRQM) \cite{Kim-2009}. 
Recently the authors in \cite{Pramanik-2013} show how to improve the fidelity of teleportation through noisy channel with the aid of 
weak measurements. In this section, we show how to improve the fidelity of the shared secret in the protocol of sequential 
quantum secret sharing under the influence of damping interactions with the help of weak measurements and post-selection. We find that 
in the case of PDC since both the qubit states $\left|0\right\rangle$ and $\left|1\right\rangle$ get affected by decoherence 
this method of reverse weak measurement and post-selection cannot be employed to improve the fidelity of the shared secret. 
But in the case of ADC, unlike PDC, $\left|0\right\rangle$ state remains unaltered and hence we find that the method of WMRQM 
can be applied to improve the fidelity of the shared secret.

In this protocol we apply weak measurements and its reversal, in two stages: one before and other after the decoherence 
acts on the system. After preparing the state represented in Eq.(~\ref{eq:inpstate}), Alice 
makes a weak measurement \cite{Kim-2009,Pramanik-2013} of strength $s_{i}$ on the $i^{th}$ qubit where $i=\{1, 3\}$. 
The weak measurement here is ensured by reducing the sensitivity of the detector. On one hand 
the detector clicks with probability $s_{i}$ if the input qubit is in state $\left|1\right\rangle _{i}$ 
and subsequently the protocol fails since the input state collapses on $\left|1\right\rangle _{i}$ in a irreversible way. 
On the other hand the detector never clicks if its in state $\left|0\right\rangle _{i}$ but it partially biases the input state towards 
$\left|0\right\rangle _{i}$ which remains unaffected by the damping interaction given by Eq.(~\ref{eq:noiserow}). 
The measurement operator corresponding to the detection and non-detection of the qubits are respectively given by,

\begin{eqnarray}
M_{q,1}=\left(\begin{array}{cc}
0 & 0\\
0 & \sqrt{s_{q}},
\end{array}\right) \label{eq:weak1}
\end{eqnarray}

\noindent which is irreversible since it doesn't possess an inverse,

\noindent and 

\begin{eqnarray}
M_{q,0} = \left(\begin{array}{cc}
1 & 0\\
0 & \sqrt{1-s_{q}}
\end{array}\right) \label{eq:weak0}
\end{eqnarray}

\noindent which is reversible as it has an inverse. Here $M_{q,0}^{\dagger}M_{q,0}+M_{q,1}^{\dagger}M_{q,1}=I$ and $q=\{1,3\}$ 
represent the qubit on which the measurement is being performed. 
Alice makes this weak measurement on 1st and 3rd qubit and sends them to Charlie and Bob respectively.

If $\rho_{in}$ be the density matrix corresponding to the pure state $\left|\Psi_{in}\right\rangle$ in Eq.(~\ref{eq:inpstate}), then 
the output state after the measurement is a mixed state and is denoted by $\rho^{WW}$

\begin{eqnarray}
\rho^{WW}=(M_{1,0}\otimes I\otimes M_{3,0})\rho_{in}(M_{1,0}^{\dagger}\otimes I^{\text{\dag}}\otimes M_{3,0}^{\dagger}). \label{eq:rowweak}
\end{eqnarray} 

For our convenience, we assume the input state parameters, $\alpha^2=k$ and $\beta^2=1-k$.
The success probability of the measurement or in other words the detector's inefficiency is given by, 
$SP_{1}=Tr\left[\rho^{WW}\right]=\frac{1}{2}(1+\bar{s})(1-\bar{k}s)$, where $\bar{s}=(1-s)$ and $\bar{k}=(1-k)$.

Alice then sends both the qubits (1 and 3) to Charlie and Bob through
the ADC where they suffer decoherence due to interaction with the
environment. Owing to this effect the mixed state $\rho^{WW}$ takes the form,

\begin{eqnarray}
\rho^{DD}=\underset{}{\sum_i}\underset{}{\sum_j}[(K_{i}\otimes I\otimes K_{j})\rho^{WW}(K_{i}^{\text{\dag}}\otimes I^{\text{\dag}}\otimes K_{j}^{\text{\dag}})] \label{eq:rownoise}
\end{eqnarray}

where $i,\: j=\left\{ \:0,\:1\right\} $ in the case of ADC.

Lastly, Bob and Charlie perform the reverse quantum measurement \cite{Kim-2009,Pramanik-2013} $N_{z,0}$ 
(corresponding to $M_{q,0}$ in Eq.(~\ref{eq:fourstatesADC}) on the qubits received from Alice after suffering decoherence. The operators for 
reverse weak measurements are given by,

\begin{eqnarray}
N_{z,0} = \left(\begin{array}{cc}
\sqrt{1-r_{z}} & 0\\
0 & 1              \label{eq:rev0}
\end{array}\right)
\end{eqnarray}

where $r$ denotes the reverse quantum measurement strength and $z=\{1,3\}$ denotes the qubit on which measurement is being performed.

The output state, after the measurement, is again a mixed state and is denoted
by $\rho^{RR}$ and is given by,

\begin{eqnarray}
\rho^{RR}=(N_{1,0}\otimes I\otimes N_{3,0})\rho^{DD}(N_{1,0}^{\dagger}\otimes I^{\text{\dag}}\otimes N_{3,0}^{\dagger}) \label{eq:rowrev}
\end{eqnarray}

In this case, the overall success probability,\\
\begin{eqnarray}
SP_{2}=\frac{1}{2}(k\bar{r}-\bar{k}\delta\bar{s})(2-(1+p)r+ \delta s)
=Tr\left[\rho^{RR}\right] \label{eq:succ}
\end{eqnarray}
where $\delta=(pr-1)$, $\bar{r}=(1-r)$, $\bar{s}=(1-s)$ and $\bar{k}=(1-k)$.

For simplicity, here we assume the measurement strengths to be uniform, $s_{1}=s_{3}=s$ and $r_{1}=r_{3}=r$. 

To reconstruct the secret at Bob's end, Alice measures her qubit
in computational basis $\{|0\left\rangle \right.$, $|1\left\rangle \right.\}$ and
Charlie measures in horizontal basis $\{|+\left\rangle \right.$, $|-\left\rangle \right.\}$. 
After these measurements Bob's state collapses to one of the
four states, $\rho_{out}^{a,c}$ , with $a\:\epsilon\:\{0,1\}$ representing
Alice's measurement outcomes and $c\:\:\epsilon\:\:\{+,-\}$, representing
Charlie's measurement outcomes. Alice and Charlie send their measurement outcome to 
Bob over classical channel. Just as before depending on these results, Bob applies the
optimal unitary transformations (same as the ones given by TABLE ~\ref{tbl:operators}), to reconstruct 
the secret. 

We illustrate two cases separately, one when the Alice's measurement outcome is $\left|0\right\rangle$ and the other when 
its $\left|1\right\rangle$, since the expression of the fidelity of the shared secret obtained in these two cases are not 
the identical, similar to the situation when no weak measurement is performed.

\textit{Case I.} When Alice's measurement outcome is $\left|0\right\rangle$, then the output density matrix, $\rho_{out}^{0,+}$, is as follows,

\begin{equation}
\left(
\begin{array}{cc}
 \frac{(-1+r) \left((-1+r) \alpha ^2+p (-1+s)^2 \beta ^2 \delta \right)}{\eta } & -\frac{j \alpha  \beta }{\eta } \\
 -\frac{j \alpha  \beta }{\eta } & \frac{(-1+p) (-1+s)^2 \beta ^2 \delta }{\eta }  \label{eq:outcome0}
\end{array}
\right)
\end{equation}

where $\delta=(pr-1)$, $\eta=(r-1)^2 \alpha^2 + (pr-1)^2 (s-1)^2 \beta^2$ and $(p-1)(r-1)(s-1)=j$.

The fidelity of the shared secret ($F_0^{WW}$), is given by, 
\begin{equation}
F_0^{WW}=\frac{k^{2}\bar{r}^{2}-\overline{k}{}^{2}\bar{s}^{2}\bar{p}\delta+k\overline{k}\bar{s}\bar{r}\left(2-\left(1+s\right)p-\bar{s}p^{2}r\right)}
{k\bar{r}^{2}+\overline{k}\bar{s}^{2}\delta^{2}} \label{eq:fidweak0}
\end{equation}
where $\delta=(pr-1)$, $\bar{r}=(1-r)$, $\bar{p}=(1-p)$, $\bar{k}=(1-k)=\beta^2$ and $\bar{s}=(1-s)$.

In this protocol, the role of forward weak measurement is to project the input state towards $\left|0\right\rangle$, which is unaffected by 
the environmental interaction. One can think of this as a map from $\left|1\right\rangle$ to $\left|0\right\rangle$, which is reversible. 
Since we have use a map from $\left|1\right\rangle$ $\rightarrow$ $\left|0\right\rangle$ at beginning so we need to again use a reverse map 
i.e., from $\left|0\right\rangle$ $\rightarrow$ $\left|1\right\rangle$ later. Moreover as the state suffers decoherence in its transit, 
we use the optimal strength of reverse quantum measurement which is a function of noise. 
Thus its crucial that the optimization should always be done on reverse quantum measurement strength ($r$) and 
not on forward one ($s$).

To achieve the objective of the protocol one needs to choose the proper strengths of the weak measurement. The optimum reverse 
measurement strength $r_{opt}$ which maximally protects the fidelity of the shared secret ($F_0^{WW}$) is obtained by maximizing 
Eq.(~\ref{eq:fidweak0}) with respect to $r$. The optimum reverse measurement strength, in this case, is given by 


\begin{eqnarray}
r_{opt}=\frac{1+(2k-1)s}{f}-\sqrt{-\frac{k\bar{p}^{2}\bar{s}^{2}}{\left(k\left(p^{2}\bar{s}^{2}-1\right)-p^{2}\bar{s}^{2}\right)f^{2}}}, \label{eq:ropt0}
\end{eqnarray}
where $f=p+2k(1-p\bar{s})-ps$ and the range of optimality is given by the following condition, 
$0<p<1$ and $0<s<1$ and $\left(\frac{-p+ps}{-2-2p+2ps}<k<\frac{-1+s}{-4+2s}\:or\:\frac{-1+s}{-4+2s}<k<1\right)$. 

Substituting this expression for $r_{opt}$, given by Eq.(~\ref{eq:ropt0}), in Eq.(~\ref{eq:fidweak0}), in place of $r$, one can get the expression to 
obtain optimal fidelity of the shared secret, $F_{opt}^{0}$. Hence, by averaging over the allowed range of input state parameter $k$ (as given above), the average optimal 
fidelity ($\overline{F}_{opt}^{0}$) is given by, 
\begin{eqnarray}
& \overline{F}_{opt}^{0} (s,\:p) = \frac{1}{8uv^{2}}\left\{ \left(8-p\bar{s}\left(p\bar{s}+2\right)\left(4-3p\bar{s}\right)\right)u\right\} \nonumber\\ 
& +\frac{1}{8uv^{2}}\left\{ 2p^{2}\bar{s}^{2}v^{2}\left(\ln \left[p\bar{s}\left(1-u+p\bar{s}\left(1+u-2\sqrt{\frac{2}{v}-1}\right)\right.\right.\right.\right.\nonumber\\ 
& \left.\left.\left.\left. +2p^{2}\bar{s}^{2}\sqrt{\frac{2}{v}-1}\right)\right]-\ln \left[\left(2-p^{2}\bar{s}^{2}-2u\right)v\right]\right)\right\}, \label{eq:avgFweak0}
\end{eqnarray}
for $0<p<1$ and $0<s<1$. Here, $u=\sqrt{1-p^{2}\bar{s}^{2}}$ and $v=1+p-ps$.
Thus, for a given value of $p$ (or decoherence channel strength), one can always choose a value of weak measurement strength ($s$) 
within the permitted range, given by the above condition, to calculate the average optimal fidelity, $\overline{F}_{opt}^{0}$.

Substituting the value of $r_{opt}$, given by Eq.(~\ref{eq:ropt0}), in Eq.(~\ref{eq:succ}) for $r$, 
one can get the required expression and calculate the corresponding average optimal success probability of 
the process ($\overline{SP}_{opt}^{0}$) by then integrating over the prescribed range of input state parameter($k$). 
The expression of $\overline{SP}_{opt}^{0}$ has been provided in the appendix of this paper. 
We observe that as $s\rightarrow1$, $\overline{SP}_{opt}^{0}$ tends to zero (becomes negligibly small) 
which coheres with results observed by the authors in \cite{Pramanik-2013}. A quick calculation shows that for $p=1$, 
Eq.(~\ref{eq:ropt0}) gives $r_{opt}=1$. Now when these values are substituted in Eq.(~\ref{eq:succ}) we get $SP_2=0$ and 
again it can be shown that $\overline{SP}_{opt}^{0}$ becomes zero for $p=1$. These results necessarily prove that for 
full damping strength ($p=1$) in the channel, the fidelity can never be improved. Moreover we find that for $p<1$, 
$\overline{F}_{opt}^{0}$ given by Eq.(~\ref{eq:avgFweak0}) increases with $s$ but remains less than 1 for $s<1$ and for $s=1$ 
the value of $\overline{F}_{opt}^{0}$ becomes undefined. Thus, for $p>0$, the value $\overline{F}_{opt}^{0}=1$ 
is impossible to obtain.

In FIG.~\ref{fig:zero}3, we show how $\overline{F}_{opt}^{0}$ varies with weak measurement strength ($s$) 
at different values of decoherence strength ($p$) and thus remains protected by WMRQM. 
Here, we simultaneously compare how the $\overline{SP}_{opt}^{0}$  decays rapidly with 
increase in weak measurement strength ($s$). We also note that $\overline{SP}_{opt}^{0}$ is inversely 
proportional to the increase in the decoherence strength ($p$). This comparison clearly illustrates that 
the increase in average optimal fidelity comes at the cost of decrease in average optimal success 
probability of the process.

\begin{widetext}
\begin{centering}

\begin{table}[H]  \label{fig:zero}
\begin{flushleft}
\textbf{For Decoherence Strength, $p=0.15$:}
\end{flushleft}
\begin{center}
\begin{tabular}{c c}
\includegraphics[height=3.8cm,width=7.2cm]{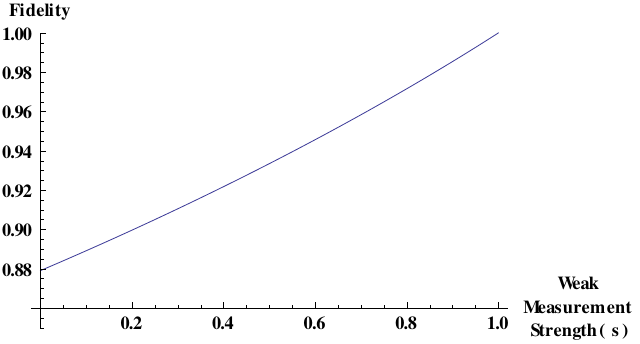} & \includegraphics[height=4cm,width=7.5cm]{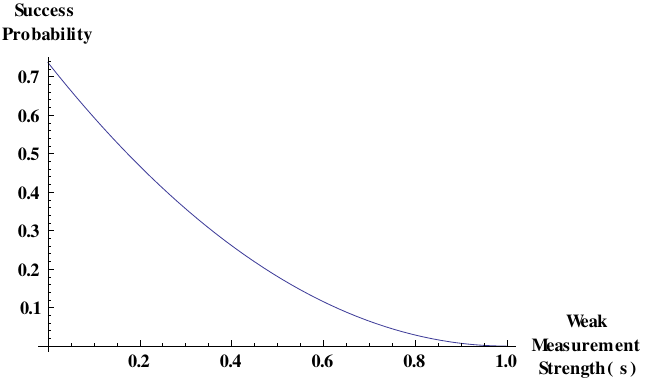}\tabularnewline
\end{tabular}\\
\end{center}
\begin{flushleft}
\textbf{For Decoherence Strength, $p=0.5$:} 
\end{flushleft}
\begin{center}
\begin{tabular}{c c}
\includegraphics[height=3.8cm,width=7.2cm]{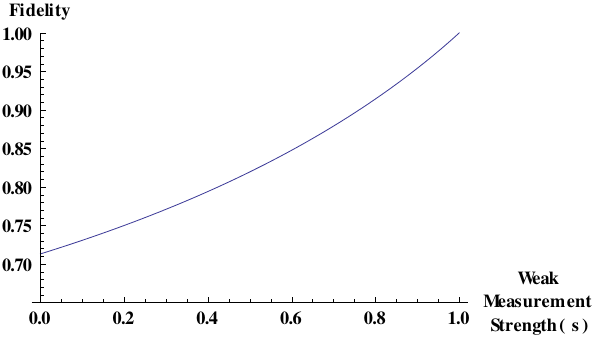} & \includegraphics[height=4cm,width=7.5cm]{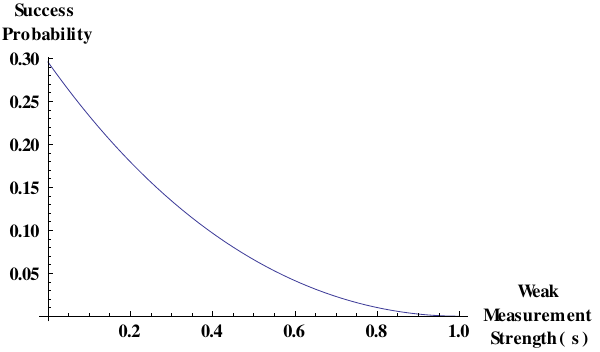}\tabularnewline
\end{tabular}\\
\end{center}
%
%
%
%
\begin{flushleft}
\textbf{For Decoherence Strength, $p=0.85$} 
\end{flushleft}
\begin{center}
\begin{tabular}{c c}
\includegraphics[height=3.8cm,width=7.2cm]{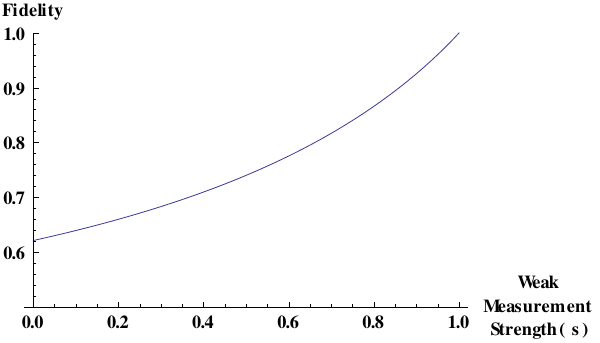} & \includegraphics[height=4cm,width=7.5cm]{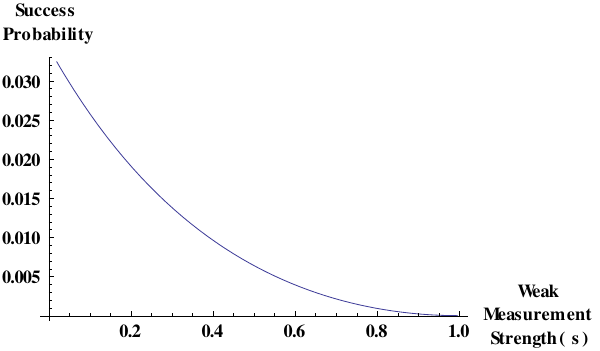}\tabularnewline
\end{tabular}
\end{center}
\begin{center}
\small FIG. 3. Variation of Average Optimal Fidelity ($\overline{F}_{opt}^{0}$) and corresponding Success Probability 
($\overline{SP}_{opt}^{0}$) with Weak Measurement Strength ($s$) for few given values of Decoherence Strength ($p$). 
\end{center}
\end{table}

\end{centering}
\end{widetext}

\textit{Case II.} When Alice's measurement outcome is $\left|1\right\rangle$, the output density matrix, $\rho_{out}^{1,+}$ after applying 
the unitary transformation $\sigma_x$, is given by,

\begin{equation}
\frac{1}{1-pr}\left(\begin{array}{cc}
\bar{p}\alpha^{2} & \bar{p}\alpha\beta\\
\bar{p}\alpha\beta & 1-pr+\bar{p}\alpha^{2}
\end{array}\right) \label{eq:rowweak1}
\end{equation}

where $\bar{p}=(1-p)$.

For our convenience, we again assume that the input state parameters, $\alpha^2=k$ and $\beta^2=1-k$. 
In this case, the final fidelity of the shared secret ($F_1^{WW}$), is given by, 
\begin{equation}
F_1^{WW}=\frac{p(k+r-kr)-1}{pr-1}. \label{eq:fidweak1}
\end{equation}

In this case, we obtain the average probability of success by first substituting $s=0$ in Eq.(~\ref{eq:succ}) and then integrating it 
over the allowed range of input state parameter ($k$) and is given by, 
\begin{eqnarray}
\overline{SP}^{1} (r,\: p)= \frac{1}{4} (r+pr-2)^2. \label{eq:avgsucc1}
\end{eqnarray}

Interestingly, in Eq.(~\ref{eq:rowweak1}) the strength of weak measurement parameter ($s$) vanishes and the reverse quantum 
measurement parameter ($r$) is the only parameter left to protect the fidelity. To achieve the objective of the protocol 
one needs to choose proper strengths of reverse measurement. The optimum reverse measurement strength which maximally 
protects the fidelity of the shared of secret ($F_1^{WW}$) is obtained putting $r=1$. With this substitution in Eq.(~\ref{eq:fidweak1}) 
the optimal fidelity becomes $1$, but the average success probability ($\overline{SP}^1$) in Eq.(~\ref{eq:avgsucc1}) corresponding to this case reduces 
to very small value. 

The average fidelity ($\overline{F}^{1}$) obtained by integrating Eq.(~\ref{eq:fidweak1}) over the allowed range of 
input state parameter ($k$) and is given by,
\begin{eqnarray}
\overline{F}^{1} (r,\: p)= \frac{p+p r-2}{2 p r-2} \label{eq:avgfidweak1}
\end{eqnarray}
for $0 \leq p<1$ and $0 \leq r<1$. 

As in Case I, here also we observe that as $r\rightarrow1$, $\overline{SP}^{1}$ becomes negligibly small. 
A quick calculation again shows that for $p=1$, the use of optimal value of reverse measurement strength ($r=1$) in 
Eq.(~\ref{eq:avgfidweak1}) gives $\overline{F}^{1}=1$ as discussed earlier, but correspondingly we also get 
$\overline{SP}^1=0$ from Eq.(~\ref{eq:avgsucc1}) with these values. Now if instead non-optimal value of $r$ is used in 
the above case, then $\overline{F}^{1}$ from Eq.(~\ref{eq:avgfidweak1}) becomes $1/2$ for all ($r<1$) which is the same 
result as obtained for $p=1$ when no WMRQM was employed. Thus, this result essentially shows that for full damping strength 
($p=1$) in the channel, the average fidelity cannot be improved and in particularly the value $\overline{F}^{1}=1$ is impossible 
to obtain. For $p<1$, we can obtain $\overline{F}^{1}=1$ using optimal reverse measurement strength of $r=1$ but with very low 
yet finite probability of success. This observation is not in terms with the results in Case I, since the $s$ parameter which 
used to lower down the success probability in Case I is absent here and for $s=1$, $\overline{F}_{opt}^{0}$ was undefined but 
here for $r=1$, $\overline{F}^{1}$ is defined and has a finite value of 1. 

To analyze what values of reverse measurement strength ($r$) will give better average success probabilities ($\overline{SP}^1$) 
as well as increase the average fidelity of the shared secret, next we plot $\overline{F}^{1}$ varying with 
strength of reverse measurement ($r$) and correspondingly compare the variation of $\overline{SP}^1$ with $r$ in FIG.~\ref{fig:one}4 
for few different values of decoherence strength ($p$). 
%
%
%
%

\begin{widetext}
\begin{centering}

\begin{table}[H] 
\begin{flushleft}
\textbf{For Decoherence Strength, $p=0.15$:}
\end{flushleft}
\begin{center}
\begin{tabular}{c c}
\includegraphics[height=3.5cm,width=7.2cm]{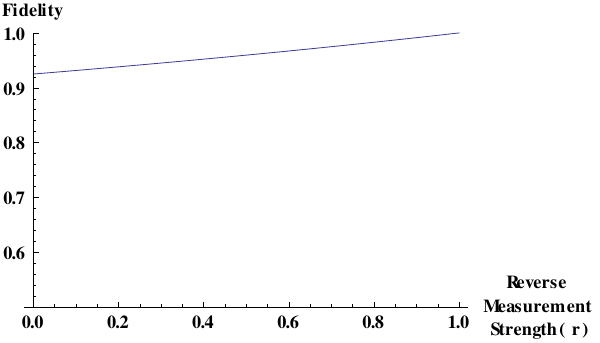} & \includegraphics[height=4cm,width=7.5cm]{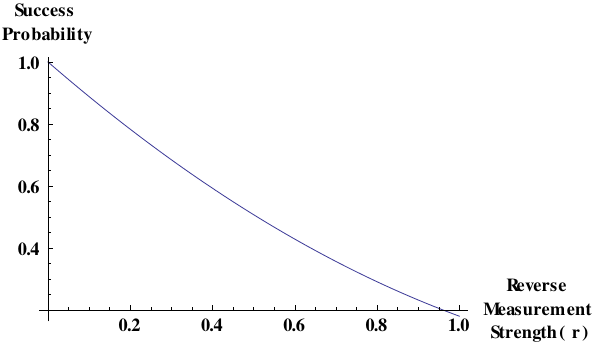}\tabularnewline
\end{tabular}\\
\end{center}
\begin{flushleft}
\textbf{For Decoherence Strength, $p=0.5$:} 
\end{flushleft}
\begin{center}
\begin{tabular}{c c}
\includegraphics[height=3.5cm,width=7.2cm]{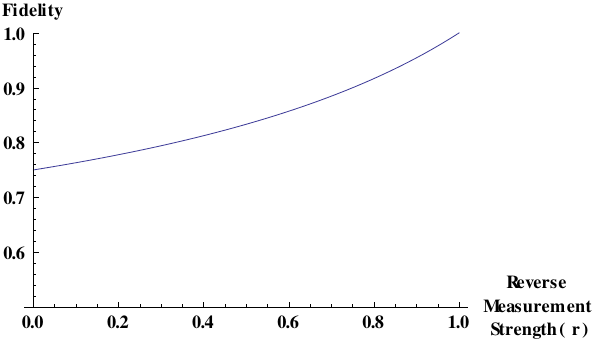} & \includegraphics[height=4cm,width=7.5cm]{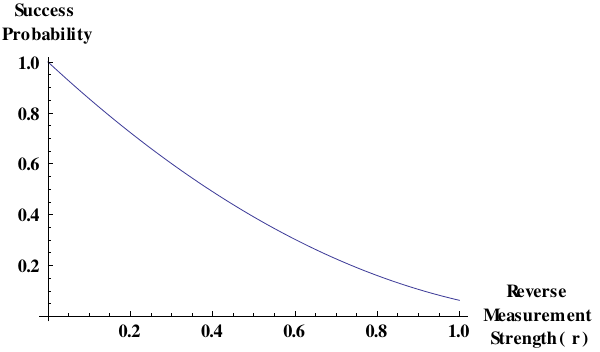}\tabularnewline
\end{tabular}\\
\end{center}
\begin{flushleft}
\textbf{For Decoherence Strength, $p=0.85$} 
\end{flushleft}
\begin{center}
\begin{tabular}{c c}
\includegraphics[height=3.5cm,width=7.2cm]{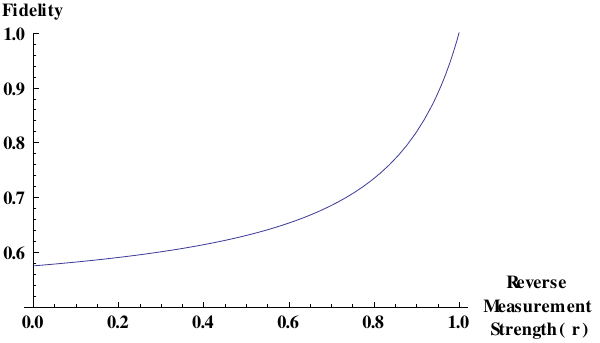} & \includegraphics[height=4cm,width=7.5cm]{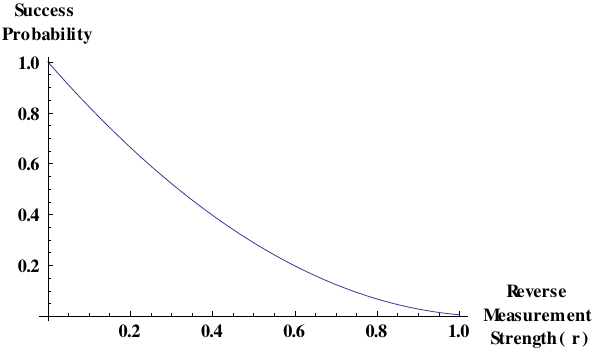}\tabularnewline
\end{tabular}
\end{center}
\begin{center}
\small FIG. 4. Variation of Average Fidelity ($\overline{F}^{1}$) and corresponding Success Probability 
($\overline{SP}^{1}$) with Reverse Measurement Strength ($r$) for few given values of Decoherence Strength ($p$).  \label{fig:one}
\end{center}
\end{table}

\end{centering}
\end{widetext}

In general for this protocol we find that in Case I the average optimal fidelity ($\overline{F}_{opt}^{0}$) stays close to 
$\frac{3}{5}$ for very high decoherence strength even at $s=0$. Again in Case II we observe that the 
parameter of weak measurement strength ($s$) vanishes in Eq.(~\ref{eq:fidweak1}) yet the reverse quantum measurement parameter ($r$) 
alone well protects fidelity from decoherence effects and with good enough success probability of the process.  
This clearly illustrates that the role of reverse quantum measurement ($r$) in protecting the fidelity of the shared 
secret is more prominent than the forward one.

Finally, in TABLE ~\ref{tbl:comparison}, we compare the improvement of average fidelities from the case when no WMRQM is employed to the ones when WMRQM 
is employed and also note the corresponding average success probabilities of obtaining such improvement at different values of decoherence channel strength 
($p$). We do this analysis corresponding to both measurement outcomes of Alice (Case I \& II). Interestingly enough we observe that in Case I 
for very low values of decoherence strength ($p$) and weak measurement strength ($s$) the method of WMRQM is not very helpful in improving of 
the average fidelity of the shared secret than in the general case when no such WMRQM is employed. However for higher values of weak measurement 
strength ($s$) and for channels not fully noisy ($p<1$) the improvement in average fidelity of the shared secret can be obtained using WMRQM 
with small enough success probability of the process. In Case II, we find that the improvement effects of this method are more prominent 
for lower values of decoherence strength as compared to Case I, than for the higher values. And we also note that the success 
probabilities of obtaining such improvements are comparatively much higher here than the former case. Its is quite evident from these values 
that there is certainly a direct trade off between the improvement of average fidelity and the corresponding success probability and 
this trade off increases with increase in the strength of weak (or reverse) measurement as well as with increase of noise in the channel.

\begin{widetext}
\begin{center}
\begin{table}[H]
\caption{\bf Comparison of trade off between the increase of Average Fidelities and corresponding Success Probabilities, 
for both the cases, for given values of Decoherence Parameter and Weak/Reverse Measurement Strength\\ 
\label{tbl:comparison}}

\begin{tabular}{|c|c|c|c|c|c|c|}
\hline 
Decoherence & Average Fidelity  & Weak or Reverse & \multicolumn{2}{c|}{Average } & \multicolumn{2}{c|}{Average}\tabularnewline
\hline 
Strength & without WMRQM & Measurement & \multicolumn{2}{c|}{Fidelity} & \multicolumn{2}{c|}{Success Probability}\tabularnewline
\hline 
$p$ & $\overline{F}_{AD}\left(p\right)$ & Strength: $s$ or $r$ & Case I: $\overline{F}_{opt}^{0}\left(s,p\right)$ & Case II: $\overline{F}^{1}\left(r,p\right)$ & Case I: $\overline{SP}_{opt}^{0}\left(s,p\right)$ & Case II: $\overline{SP}^{1}\left(r,p\right)$\tabularnewline
\hline 
\hline 
\multirow{4}{*}{0.15} & \multirow{4}{*}{0.925} & 0.25 & 0.91 & 0.94 & 0.41 & 0.733\tabularnewline
\cline{3-7} 
 &  & 0.45 & 0.93 & 0.96 & 0.219 & 0.549\tabularnewline
\cline{3-7} 
 &  & 0.65 & 0.95 & 0.97 & 0.088 & 0.392\tabularnewline
\cline{3-7} 
 &  & 0.85 & 0.98 & 0.99 & 0.016 & 0.261\tabularnewline
\hline 
\multirow{4}{*}{0.5} & \multirow{4}{*}{0.75} & 0.25 & 0.76 & 0.79 & 0.156 & 0.66\tabularnewline
\cline{3-7} 
 &  & 0.45 & 0.81 & 0.82 & 0.08 & 0.439\tabularnewline
\cline{3-7} 
 &  & 0.65 & 0.86 & 0.87 & 0.031 & 0.263\tabularnewline
\cline{3-7} 
 &  & 0.85 & 0.93 & 0.93 & 0.006 & 0.131\tabularnewline
\hline 
\multirow{4}{*}{0.85} & \multirow{4}{*}{0.575} & 0.25 & 0.67 & 0.6 & 0.016 & 0.591\tabularnewline
\cline{3-7} 
 &  & 0.45 & 0.72 & 0.62 & 0.008 & 0.341\tabularnewline
\cline{3-7} 
 &  & 0.65 & 0.8 & 0.67 & 0.003 & 0.159\tabularnewline
\cline{3-7} 
 &  & 0.85 & 0.89 & 0.77 & 0.001 & 0.046\tabularnewline
\hline
\end{tabular}
\begin{flushleft}
\textbf{Note: Here Case I \& II refers to the two cases considered by us corresponding to the two measurement outcomes of Alice, 
$\left|0\right\rangle$ and $\left|1\right\rangle$ respectively.}
\end{flushleft}

\end{table}
\end{center}
\end{widetext}

\section{Conclusions}
In a nutshell, in this work we have given a protocol for sequential secret sharing of quantum information. 
Initially we have given the protocol for three parties and then we have generalized it for $(n>3)$ parties.  We also explicitly show how our protocol is more efficient in terms of resource requirements, both classical and quantum, over the existing protocols for the purspose sharing secrets in a sequential (deferred) manner.
Not only that,  we have further considered  a much more realistic situation, where we have shared the qubits through 
two kinds of noisy channels, namely the phase damping channel (PDC) and the amplitude damping channel (ADC). When we carry out the sequential secret sharing in the presence of noise we observe that the fidelity of secret sharing at the $k^{th}$ iteration is independent of the effect of noise at the $(k-1)^{th}$ iteration. 
In case of PDC, the average fidelity ranges from $0.67$ to $1$, whereas in case of ADC we have observed that the average fidelity 
ranges from $0.5$ to $1$. In order to enhance the fidelity of secret sharing in ADC, we have employed the technique of 
weak measurement and its reversal. The improvement of the average fidelity comes at the cost of drop in success probability of the process. 
By applying WMRQM, for higher values of weak measurement strength ($s$) and for channels not fully noisy ($p<1$) we are able to improve 
the average fidelity of the shared secret with small enough success probability of the process. We have also shown that for $p=1$, the 
average fidelity of secret sharing can never be improved and more particularly the value of the average fidelity equal to 1 is impossible to obtain.

\noindent\textit{ Acknowledgment:} Authors gratefully acknowledge Dr. K. Srinathan , Mr. T. Pramanik, Mr. V. Chiranjeevi, Mr. S. Sazim 
and Mr. N. Ganguly for giving valuable advises in the improvement of this work.

\begin{widetext}
\begin{center}
 \textbf{APPENDIX}
\end{center}
\begin{eqnarray*}
& \overline{SP}_{opt}^{0}(s,\:p) = (p-1)^{2} px^{2} \sqrt{hpx}\: \left\{ 2 - 2x - p \left( p^{2} - 1 - 2p + x + px - p^{2} x \right)  + s \{-2 + p (2+ 6p + p^{2}x - 4px - x - 4p^{2} \right. \nonumber\\
& + s ( 6p^{2} - 1 - 6p + 2px - 3p^{2}x + s \left( 2p - 4p^{2} + sp^{2} + p^{2}x \right) ) ) \} - 2x \sqrt{pz} \{ 8+ p (4(5s+3) -5p + 10p^2  \nonumber\\
& - 4p^3 + 2p^4 + p^5 + s ( - 10p - 15p^2 + 11p^3 - 9p^4 - 10p^5 + s ( p^3 + 6p^4 + 5p^5 - s(p^4+p^5) )))))\} + 8p (1+p) \nonumber\\
&  x (px(px-1))^\frac{3}{2} (x+y) ArcTanh\:[1+px] + (-\sqrt{-p^5 y} - 2 \sqrt{-p^7 y} + \sqrt{-p^9 y} + p^3 (3+s (-13+3 (7-5 s) s)) \nonumber\\
&  \sqrt{p z} + 5 p^5 \left(1+s^2\right)^2 x \sqrt{p z} + 3 \sqrt{p^9 z} + 4 \sqrt{p^{11} z} + 2 p x^2 (\sqrt{-p y} - 3 \sqrt{p z}) + p^2 (-7 \sqrt{p z}+s (15 \sqrt{p z}  \nonumber\\ 
& + s (-7 \sqrt{-p y} - 9 \sqrt{p z} + s \sqrt{p z} ) ) )-p^4 s (4 \sqrt{-p y} + 11 \sqrt{p z} + s (-14 \sqrt{p z} + s (4 \sqrt{-p y} + 6 \sqrt{p z} + s \sqrt{p z} ) )) \nonumber\\
& + s (5 (\sqrt{-p^5 y} + \sqrt{-p^7 y} ) + s (-3 \sqrt{-p^7 y} + 6 \sqrt{-p^9 y} + s (3 \sqrt{-p^5 y} - \sqrt{-p^7 y} + s (\sqrt{-p^7 y} + \sqrt{-p^9 y} + 4 \sqrt{p^7 z}  \nonumber\\
& + s (\sqrt{p^9 z} - 4 \sqrt{p^{11} z}) ) ) )) ) Log\: [4] + 256p^{7} (1+p)^{4} x^{7}y(2+px)^{3} (3px-1)(1+3px)(px(1+px)-2) z^{5} \nonumber\\
& + (-2 p \sqrt{p z} - 5 \sqrt{p^5 z} - 3 \sqrt{p^7 z} + \sqrt{p^9 z} + \sqrt{p^{11} z} - s (-3 (\sqrt{p^5 z} + \sqrt{p^7 z} - \sqrt{p^9 z} - \sqrt{p^{11} z}) + (s-3) s \nonumber\\
& (\sqrt{p^9 z} + \sqrt{p^{11} z}))) \:Log\:[p-p s]\: Log\:[\frac{2}{1+p-p s}]\: Log\:[1+p-p s] \:Log\:[1+\frac{1}{-1+p x}]\:Log\:[p z] + 4 (1+p) (x+y) (-p y)^{\frac{3}{2}}  \nonumber\\
& (Log\:[-2 i (-1+\sqrt{1-p^2 x^2})] - 4 x^3 (p^5 x^2 \sqrt{p z}-2 \sqrt{p^5 z} - 3 \sqrt{p^7 z} + s (\sqrt{p^7 z}-\sqrt{p^9 z}+s\sqrt{p^9 z}))\nonumber\\
& Log\:[2+p x]\: Log\:[i (-\sqrt{-p y}+\sqrt{p z})])))\} \left/\:(8 (1+p) (1+p (-1+s))^4 (2+p (-1+s)) (p (-1+p (-1+s)) (-1+s))^{\frac{3}{2}})\right.
\end{eqnarray*}
where $x=s-1$, $z=x(px-1)$, $h=(p(s-1)-1)(1+p^2(s-3)(s-1))$ and $y=x(px+1)$.\\
\end{widetext}

\begin{thebibliography}{1}

\bibitem{einstein}A. Einstein, B. Podolsky and N. Rosen, Phys. Rev. \textbf{47}, 777 (1935).

\bibitem{bennett2} C. H. Bennett, G. Brassard, C. Crepeau, R. Jozsa, A. Peres, W. K. Wootters, Phys. Rev. Lett. \textbf{70}, 1895 (1993); 
D. Bouwmeester, J.W. Pan, K. Mattle, M. Eibl, H. Weinfurter and A. Zeilinger, Nature \textbf{390}, 575 (1997).

\bibitem{gisin} N. Gisin, G. Ribordy, W. Tittel, and H.Zbinden, Rev. Mod. Phys. \textbf{74}, 145 (2002).

\bibitem{hillery} M. Hillery, V. Buzek, and A. Berthiaume, Phys. Rev. A \textbf{59}, 1829 (1999); 
R. Cleve, D. Gottesman, and H-K. Lo, Phys. Rev. Lett. \textbf{83}, 648 (1999).

\bibitem{cleve99} R. Cleve et.al, Phys.Rev.Lett. \textbf{83} 648-651 (1999). 

\bibitem{karlsson} A. Karlsson, M. Koashi, and N. Imoto, Phys. Rev. A \textbf{59}, 162 (1999).

\bibitem{bandyopadhyay} S. Bandyopadhyay, Phys. Rev. A \textbf{62}, 012308 (2000).

\bibitem{bagherinezhad} S. Bagherinezhad, and V. Karimipour, Phys. Rev. A \textbf{67}, 044302 (2003).

\bibitem{lance}A. M. Lance, T. Symul, W. P. Bowen, B. C. Sanders, and P. K. Lam, Phys. Rev. Lett. \textbf{92}, 177903 (2004).

\bibitem{Koashi-1999} M. Koashi and M. Ueda, Phys. Rev. Lett. \textbf{82}, 2598 (1999).

\bibitem{Kim-2009} Y.S. Kim, Y.W. Cho, Y.S. Ra, Y.H. Kim, Opt. Express \textbf{17}, 11978 (2009).

\bibitem{Lee-2011} J.C. Lee, Y.C. Jeong, Y.S. Kim, and Y.H. Kim Opt. Express \textbf{19}, 16309 (2011).

\bibitem{Kim-2012} Y.S. Kim, Y.C. Lee, O. Kwon and Y.H. Kim, Nat. Phys. \textbf{8}, 117 (2012).

\bibitem{Pramanik-2013} T. Pramanik, and A.S. Majumdar, Phys. Lett. A \textbf{377}, 3209-3215 (2013).

\bibitem{Nielsen-2002} M.A. Nielsen and I.L. Chuang, “Quantum Computation and Quantum Information”, Cambridge University Press (2002).

\bibitem{Aharonov-1988} Y. Aharonov, Z.D. Albert, and L. Vaidman, Phys. Rev. Lett. \textbf{60}, 1351-1354 (1988); 
I.M. Duck, P.M. Stevenson, and E.C.G. Sudarshan, Phys. Rev. D \textbf{40} 2112 (1989).

\bibitem{gordon} G. Gordon, and G. Rigolin, Phys. Rev. A

\bibitem{zheng} S. B. Zheng, Phys. Rev. A \textbf{74}, 054303 (2006).

\bibitem{markham} A. Keet, B. Fortescue, D. Markham, B. C. Sanders, Phys. Rev. A \textbf{82}, 062315 (2010) 

\bibitem{markham08} D. Markham, B.C. Sanders, Physical Review A \textbf{78}, 042309 (2008). 

\bibitem{li} Q. Li, W. H. Chan, and D-Y Long, Phys. Rev. A \textbf{82},
022303 (2010).

\bibitem{satya} S. Adhikari, Quantum secret sharing with two qubit bipartite mixed states, arXiv:1011.2868.

\bibitem{indranil} S.Adhikari, I. Chakrabarty, P. Agrawal, Quantum Information and Computation, \textbf{12}, 0253 (2012).  

\bibitem{tittel} W. Tittel, H. Zbinden, and N. Gisin, Phys. Rev. A \textbf{63},
042301 (2001).

\bibitem{schmid} C. Schmid, P. Trojek, M. Bourennane, C. Kurtsiefer, M. Zukowski,
and H. Weinfurter, Phys. Rev. Lett. \textbf{95}, 230505 (2005).

\bibitem{schmid1} C. Schmid, P. Trojek, S. Gaertner, M. Bourennane, C. Kurtsiefer, M. Zukowski, 
and H. Weinfurter, Fortschritte der Physik \textbf{54}, 831 (2006).

\bibitem{bogdanski} J. Bogdanski, N. Rafiei, and M. Bourennane, Phys. Rev. A \textbf{78}, 062307 (2008).

\bibitem{sazim-2013} S.k. Sazim, C. Vanarasa, I. Chakrabarty, and K. Srinathan, arXiv:1311.5378.

\bibitem{sazim-2012} S.k. Sazim and I. Chakrabarty, Eur. Phys. J. D \textbf{67}, 8 (2013) 174.

\bibitem{massar-1995} S. Massar and S. Popescu, Phys. Rev. Lett. \textbf{74}, 1259 (1995).

\end{thebibliography}
\end{document}